\documentclass[12pt,preprint]{aastex}

\def\etal{{\it et al.~}}

\shorttitle{Variability of Sagittarius A*}
\shortauthors{Zhao \etal}

\begin{document}


\title{Variability of Sagittarius A* - Flares at 1 Millimeter}


\author{Jun-Hui Zhao\altaffilmark 1, K. H. Young\altaffilmark 1, 
R. M. Herrnstein\altaffilmark 1, P. T. P. Ho\altaffilmark 1, 
T. Tsutsumi\altaffilmark 1,
K. Y. Lo\altaffilmark 2,
W. M. Goss\altaffilmark 3, and
G. C. Bower\altaffilmark 4}

\altaffiltext 1
{Harvard-Smithsonian CfA, 60 Garden St, MS 78, Cambridge, MA 02138;
jzhao@cfa.harvard.edu, kyoung@cfa.harvard.edu, rmcgary@cfa.harvard.edu,
pho@cfa.harvard.edu, ttsutsumi@cfa.harvard.edu}

\altaffiltext 2
{Academia Sinica Institute of Astronomy \& Astrophysics, Taipei,
Taiwan;flo@nrao.edu}

\altaffiltext 3
{NRAO-AOC, P. O. Box 0, Socorro, NM 87801; mgoss@aoc.nrao.edu}

\altaffiltext 4 {University of California, Berkeley, CA94720; gbower@astron.berkeley.edu}


\begin{abstract}

We report the results from recent observations of Sgr~A*  
at short-/sub-millimeter
wavelengths made with the partially finished Sub-Millimeter Array 
(SMA) on Mauna Kea.
A total of 25 epochs of observations were 
carried out  over the past 15 months in 2001 March to 2002 May.
 Noticeable variations in flux density
at 1.3 mm were observed showing three ``flares''.
The SMA observations suggest that Sgr~A*
 highly increases towards  submillimeter wavelengths during a flare 
suggesting the presence of  a break wavelength in spectral index 
around 3 mm.
A cross-correlation of the SMA data at
1 mm  with the VLA data at 1 cm
show a global delay of ${\rm t_{delay}> 3 d}$,  
suggesting that sub-millimeter wavelengths
tend to peak first. 
Only marginal 
day-to-day variations in flux density (2-3$\sigma$) have been detected at 1.3 mm.
No significant  flares on  a short time scale ($\sim1$ hr) have been
observed at 1.3 mm. We also failed to detect significant
periodic signals at a level of 5$\%$ (3$\sigma$)
from Sgr A* in a periodic searching window
ranging from 10 min to 2.5 hr.
The flares observed at the wavelengths
between short-centimeter 
and sub-millimeter might be a result of collective mass
ejections associated with X-ray flares that originate from the inner region
of the accretion disk near the supermassive black hole.

\end{abstract}


\keywords{Galaxy:center --- accretion, accretion disks ---
galaxies:active --- radio continuum: galaxies ---
black hole physics}


\section{Introduction}
Sgr~A*, a compact radio source,   is believed
to be associated with the supermassive black hole at the Galactic center
(Eckart \etal, 2002; Ghez \etal, 2000).
The inferred bolometric luminosity (L$\sim10^{-8.5}$ L$_{\rm Edd}$)  
is far below the Eddington luminosity for
the black hole mass of 
$\sim$2.6$\times10^6$ M$_\odot$. Sgr~A* represents an
extremely dim  galactic nucleus. The low luminosity  of Sgr A*
is explained by the low efficiency radiative 
advection dominated accretion flow
(ADAF) model (Narayan \etal~1998). 
However, the ADAF alone 
 can not explain the
detailed activities, such as the radio outbursts and the
X-ray flares.
The recent work by Yuan \etal~(2002a) 
provides an alternative  model to link
a jet outflow (Falcke \etal~1993) with the ADAF. This new model appears to be able to fit 
reasonably well  the overall spectrum of Sgr~A* from radio, sub-millimeter,
IR to the X-ray. 

The apparent structure at radio wavelengths longer than  3 mm 
appears to be  mainly  dominated by 
the scattering effects due to  the ISM. 
At 3 mm, the scattering effects are finally small enough
that the intrinsic source size is estimated 
to be  less than 0.27 mas 
(Doeleman \etal~2001),
or about 40 times the Schwarzchild radius (${\rm R_{sc}}$ hereafter). 
A promising way to explore this
extremely compact source
is to monitor the variations of the emitted flux density 
from radio to
X-ray. 

The variations in radio flux density of Sgr~A* have been known for two
decades (Brown \& Lo 1982). The nature of the radio variability has not been
well  understood. At the long wavelengths, the flux density of Sgr~A* might
be modulated by the scintillation due to the turbulence in the ISM (Zhao
\etal~1989). 
The radio light curves observed with the VLA at wavelengths from 20 to 1.3 cm
during the period of 1990-1993 suggest
that the fractional amplitude variations increased towards short wavelengths
and that the rate of radio flares appeared to be about three per year
(Zhao {\it et al.} 1992; and Zhao \& Goss 1993).
The typical time scale of these radio flares is about
a month.  The observed large amplitude
variations in flux densities at 3 mm (Wright \& Backer 1993;
Tsuboi \etal~1999) are
consistent with the wavelength-dependence of the  variability 
as observed at centimeter wavelengths.

The presence of a 106 day cycle in the radio variability of Sgr A$^*$
was suggested from an analysis of data observed with the VLA 
in the period of 1977-1999 (Zhao \etal~2001). 
The periodic oscillation at a period around  100 days
appears to persist in the densely sampled light curves obtained 
with the VLA at 2, 1.3 and 0.7 cm over the 
past two years (Zhao \etal~2002). The period of the fluctuation
cycle appears to increase
to $\sim$ 130 days  (Bower \etal~2002a; Zhao \etal~2002). 
In addition, a  longer periodic fluctuation 
 feature with a period 2.4-2.5 times the short one has also been seen
(Zhao \etal~2002). Similar periodic fluctuation in flux density 
was also observed in the GBI monitoring data (Falcke, 1999).

Observations of Sgr A* at  sub-millimeter can
penetrate into the deep region of this intriguing source. 
In this letter, we report  results obtained from a 
 monitoring program at 1 mm, with the 
partially finished Sub-Millimeter Array (hereafter SMA; Moran 1998).

\section{Observations \& Data Reduction}

Observations of Sgr~A* at 1.3 mm and 0.87 mm 
were made using the partially completed 
SMA with three or four antennas and  baselines ranging
from 7 to 55 kilo wavelengths at 1.3 mm.
A total of 24 epochs of observations at 1.3 mm were carried out in 
reasonably good weather conditions (the sky 
opacity $\tau_{\rm zenith}<$
0.3 at 1.3 mm). One observation was made at 0.87 mm
on 2001 March 22 with $\tau_{\rm zenith}\sim$0.5 at 0.87 mm.
 The observations were carried
out with a total bandwidth of 328 MHz for each sideband. 
A typical system temperature is 200-300 K at 1.3 mm.
A typical r.m.s. noise of $\sim$ 20 mJy was  achieved
from an observation with four antennas for a typical on-source
integration time of 2 hrs. 
In each epoch of observation, we interleaved Sgr A* with 
Sgr B2(N), a compact ($<5$\arcsec~in size) nearby (${\rm \sim1^o}$) HII region, and two nearby QSOs,
OV236 (${\rm \sim22^o}$) and NRAO 530 (${\rm \sim16^o}$). 
The flux density scale in each observation
was determined by observing a compact planet (Neptune ($<2.5$\arcsec~in diameter)
 and Uranus ($<4$\arcsec)).
Sgr~B2(N) ($\sim$50 Jy at 1.3 mm) was used to monitor
the stability of the telescope during the observations,
for example the effect due to the possible telescope pointing drift.

Further calibration was done by observing two QSOs, OV236 and NRAO 530.
We fitted the secular variations of each calibrator
with polynomials. Any correlated offsets in flux density 
from the secular variations of the two QSOs are considered as 
systematic offsets, such as telescope pointing error. 
The residual gain correction determined from the offsets
are applied to the Sgr A* data in order to minimize the
the systematic errors. The final uncertainty of $\sim10\%$ in the flux
density calibration is assessed by calculating the standard deviations of
the residual offsets from the calibrators. However, 
the variation of  a possible linear
polarization as a function of parallactic angle was not
corrected. A fractional linear
polarization of $\sim$7\% from Sgr A* has been detected
(Bower {\it et al.},  2002b). The error due to this effect is less
significant and is embedded in the final uncertainty.

In addition, Sgr~A* is embedded in the complex, extended source Sgr A West. 
Fig. 1 shows the image of Sgr A* and its vicinity observed
with the SMA at 1.3 mm with a beam of 7.4\arcsec$\times$2.3\arcsec
(P.A. =7\arcdeg). At this wavelength and this angular resolution,
Sgr A* is about 10 times brighter than the surrounding components.
By examining  the visibilities as a function of baseline lengths,
 we find that,
 for baselines $\sim$ 20k$\lambda$ or longer, Sgr~A* is the dominant
source and the confusing flux density at 1 mm 
from the
surrounding free-free and dust emission is less than 0.3 Jy.
The flux density measurements were made  in both the visibility and imaging
domains. In the visibility domain, we measured the amplitude of the baselines 
of $\sim$20 k$\lambda$ and longer. The flux density was double checked
by constructing images with the self-calibrated (phase-only) visibility
data. The measurements of the point source flux density were
done by deconvolving the telescope beam and taking out the contribution from
surrounding components  
using IMFIT in AIPS.
The flux densities derived from the two domains
are in an agreement within 10\% of the mean value.
This additional 10\% level of the uncertainty in the 
measurements  is mainly due
to the confusion from the surrounding emission 
and is added to the final
error assessment. The error bars of Sgr A* are
derived from the quadrature addition
of the uncertainty in flux density calibration and
the uncertainty due to the confusion. 
The typical uncertainty at 1 mm is in a range of 10 to 20\%.

\section{Results}

\subsection{Light Curve and Flares}
Fig. 2 shows the SMA light curve at 1.3 mm
suggesting that Sgr~A* varies significantly.
A few ``flares'' were observed from Sgr A*
 while the calibrators show  secular variations 
with opposite drifts in flux density over the past year (
Fig. 2a).  
Three ``flares'' were observed over a 1-year period.
Both  the 2001-March and 2002-February flares 
(Flare 1 and Flare 3 as marked in Fig. 2b)
were partially observed in  their decreasing phase.
The 2001-July flare (Flare 2) was observed covering an entire
cycle
from its inception to a slow decrease back to its steady state value.
Flare 1, 
 which started from 4.1$\pm$0.5 Jy
after an unobserved peak and decreased to 1.1 $\pm$ 0.15 Jy 
within less than three months,
appeared to be relatively stronger than others.
The rising time for
Flare 2 was about 2-3 weeks, reaching a peak of
$3.2\pm0.3$ Jy on 2001 July 10. 
Then, a slow decrease lasted about 40-50d.
We were not able to observe Sgr~A* for the next three months due to the
proximity to the Sun.
The monitoring program was resumed in 2002 February.
A tail of a possible flare (Flare 3) was  observed in early 2002. 

\subsection{Day-to-Day Variability and Non-detection Limit on Intra-day Variability}

Based on the sparse data, marginal day-to-day variations at a level of 
2-3$\sigma$  (or 20-30\%) were observed 
during Flare 1 and Flare 2 as well as in later May 2002. 

Intra-day variations on short time scales were searched
based on the 24 epochs of observations at 1.3 mm.
We averaged the data in a 5min bin and checked the visibility
plots for each baseline. We also averaged all baseline together
and checked intra-day light curves.
No
evidence for significant variations on a time scale of $\sim$1 hr
has been found,
{\it i.e.} a variability quantity ${{\left[S_{max}-S_{min}\right] 
\over \left[S_{max}+S_{min} \right]}
< 20\%}$, where ${\rm S_{max}}$ and ${\rm S_{min}}$
are the maximum and minimum flux densities at 1.3 mm, respectively, 
 in a single observing track of 6 hrs or less. 

We also searched for periodic signals in  a period ranging between
10 min to 2.5 hrs based on the observations of 5.5 hr
on May 29, 2002. 
There is a possible  
oscillation signal with a frequency of 1.1$\times10^{-4}$ Hz
(or 2.5 hr in period) on an E-W baseline. This oscillation signal can be well
modeled as a structure due to interference between Sgr A* and  surrounding
components.
Combining all baselines, a $3\sigma$ non-detection limit of  
 a periodic signal from SgrA*  at a level of$\sim$5\% can be inferred.

\subsection{Spectrum during A ``Flare''}

We also observed Sgr~A* at 0.87 mm  with the SMA on 2001 March 22.
 We observed Sgr~A* 
(S$_{\rm 0.87mm}=6.7\pm1.5$ Jy) at the sub-millimeter band using the
three-element array of the partially completed SMA.
Fig. 3a shows a spectrum derived from the mean flux density
determined from the multiple observations within
two weeks of
the peak of Flare 1. The error bars were derived from the
quadrature addition of the standard deviation and the maximum
error in individual measurements. 
The spectral index $\alpha$ (S$_\nu\propto\nu^\alpha$)
appears to be 0.1$\pm$0.1 at 100 GHz and below, and 1.5$^{+1.0}_{-1.1}$ 
between 232 and 345 GHz, suggesting a break frequency in spectral index
of $\sim$100 GHz or higher.  A flux density excess towards  sub-millimeter
wavelengths has been observed (Zylka, Mezger, \& Lesch 1992;
Serabyn \etal~1997; Falcke \etal~1998). 

The overall spectrum can  fit two power-law components, {\it i.e.}
S$_\nu$ = S$_1(\nu/\nu_1)^{\alpha_1}$ + S$_2(\nu/\nu_2)^{\alpha_2}$. 
Three sets of combination of  ${\alpha_1}$
and ${\alpha_2}$ are used in the fitting. First, for ${\alpha_1}=0.0$
and ${\alpha_2}=2$ (dashed lines in Fig. 3), the sub-millimeter
component corresponds to  the thermal synchrotron emission either
arising from the inner region of the accretion disk ({\it e.g.}
Liu \& Melia 2002) or  produced
from a jet-nozzle (Falcke \& Markoff 2000). Second, for ${\alpha_1}=0.1$
and ${\alpha_2}=2.5$ (solid lines in Fig. 3),
the spectral index of 2.5 suggests that a homogeneous opaque, non-thermal
synchrotron source might be
present in the inner region of the accretion flow.
Such a model appears to be plausible
if one considers the non-thermal synchrotron particles
 to be accelerated inside
the compact source, perhaps within a jet nozzle as has been proposed
for the case of NGC 4258 (Yuan {\it et. al.} 2002b). Finally,
if the low frequency  component has an exponential
cut-off, {\it i.e.} ${\rm \sim S_1(\nu/\nu_1)^{0.25} exp(-\nu/\nu_0)}$,
at $\nu_0 \sim 75$ GHz, a smaller value of $\alpha_2\sim1.5$ (dash-dotted lines
in Fig. 3)
 for the
sub-millimeter component is also consistent
with a spectrum produced from the ADAF in which a gradient of ${\rm T_e}$
 depresses the rising part of the spectrum (Narayan {\it
et. al.}, 1998). The observed spectrum suggests an opaque nature
of the sub-millimeter component at 1.3 and perhaps 0.87 mm. Observations
at the shorter sub-millimeter wavelengths appear to be
critical to differentiate between the models.
The spectrum in a minimum state is also shown (Fig. 3b). The excess
at 1.3 mm appears to be less significant.

\subsection{Correlation with The VLA Data}

The SMA data at 1 mm appears to show a correlation with
the light curves observed with the VLA. 
The SMA light curve shows 
three ``flares''  from  Sgr~A* in the past year
from 2001 March to 2002 May. 
The variation in flux density that we observed in the SMA 
light curve suggests that flares are 
constantly occurring in this source.

 A quantitative analysis 
of cross-correlation  properties  between the light curves at 1.3 mm 
and 1.3 cm has been carried out. Due to the sparse data sampling, large 
uncertainty remains in the cross-correlation analysis for individual flares
(Zhao 2002).  
However, a global delay between the 1.3 mm and 1.3 cm
light curves can be searched for using the z-transformed discrete 
correlation function (ZDCF) without interpolating in the temporal domain
(Alexander 1997).  With no prior models assumed,
the ZDCF is a reliable and efficient method to search for a delay.
Fig.~4 shows the ZDCF  
between  the SMA and VLA light curves at 1.3 mm and 1.3 cm.
The peak in the correlation function corresponds to a delay of 4$^{+2}_{-1}$ d.
A noticeable asymmetric shape of the ZDCF near the zero lag 
shows a significant excess in cross-correlated power towards
the positive lags,
suggesting that the true global delay is
${\rm t_{delay}>3 d}$.  The ZDCF indicates that 
flares at 1.3 mm starts first.

In addition, the strong X-ray flare with a time scale of $\sim$1 hr,
 observed by 
Baganoff \etal~(2001), occurred about 10 days earlier than a radio peak
observed in all three VLA monitoring  bands. During 
Flare 2 (2001 July), Chandra observed Sgr~A* on 2001 July 14,
a few days past the 1 mm peak but no X-ray flares were observed. 
The X-ray flux level was consistent with
that of a quiescent state (Baganoff 2001, private communication).

\section{Discussion \& Summary}
The SMA observations have shown that Sgr~A* varies significantly at 
1 mm
during the course of SMA monitoring in 2001 March to 2002 May.
The derived lags from a cross-correlation analysis 
appear to be  good evidence that the
flaring occurs from the inside out starting from short wavelengths
and then continuing to longer wavelengths.
In the  Jet-ADAF model connecting a jet outflow with the ADAF
(Yuan \etal~2002a),  the sub-millimeter excess
is thought to arise
from a sum of the emission from both the ADAF 
and the jet nozzle.

On the other hand, from the  observations with the SMA and 
the VLA, further constraints
on the models can be derived.
Taking a global  delay
time of ${\rm t_{delay} >3 d}$ and the source size of 40 R$_{\rm sc}$, an expansion
velocity, v$_{\rm exp}\sim$1200 km s$^{-1}$ or $<$ 0.004 c, 
is inferred. The expansion velocity appears to be far below
the escape velocity of 0.1 c at r $\sim$ 40 R$_{\rm sc}$. 
The bulk kinetic energy associated with the flares
appeared to be too small to power a noticeable  collimated jet in Sgr~A*. 
In addition, the break in the spectral index at $\sim$3 mm also
indicates that a large fraction of the 
flaring plasma might well be confined within
the characteristic radius at 3 mm. However, the possible outflow
tries to expand to a larger scale. The data presented here
 does not  exclude the possibility of a strong flare which can lead
  to an observable jet-like structure. The inferred
  small expansion velocity may imply that other processes 
  contribute to the 
  transport of high energy
  particles, {\it e.g.} diffusion and convection 
   may also play a role in powering Sgr A*
  at lower radio frequencies.

The time scale (weeks) of 1 mm flares differs from the
time scale (1 hr) of the X-ray flare  (Baganoff, 2001). 
The  lack
of strong flares on a short time scale at 1 mm places a critical constraint
on the models of the inverse Compton scattering as 
has been proposed for the short duration X-ray flares
(Falcke \& Markoff 2000;
Markoff {\it et. al.}, 2001; Liu \& Melia, 2002 ). 
Considering
the opaque nature of the sub-millimeter
component at 1.3 mm, the X-ray flares could
hide at 1.3 mm due to a self-absorption. 
Alternatively,
the flares at sub-millimeter wavelengths might be a result of 
collective 
mass ejections associated with the  X-ray flares that originate
from the inner region of the accretion flows near the event
horizon of the supermassive black
hole at the Galactic center.

\acknowledgments
We would like to thank the SMA staff from both the SAO and the ASIAA for
supporting this monitoring program. We are grateful to Jim Moran and Heino
Falcke (the referee) for their valuable comments. JHZ thanks
Irwin Shapiro for his initial suggestion of the SMA monitoring program.
The Very Large Array (VLA) is operated by the National Radio Astronomy 
Observatory
(NRAO). The NRAO is a facility of the National Science Foundation operated 
under
cooperative
 agreement by Associated Universities, Inc.





\clearpage



\figcaption[fig1.ps]{ A pseudo-color image of Sgr~A* (blue) and its vicinity
(red) made  from 3.5 hr observations 
using the four elements of the partially finished SMA on 2002 May 23.
The total integration time  on  the source  is about 1.5 hr. The r.m.s.
 noise is $\sim$30 mJy/beam. The FWHM beam is 
 7.4\arcsec$\times$2.3\arcsec
(P.A. = 7 \arcdeg).  The data reduction was done in the AIPS environment following a 
procedure described in a SMA Technical Memorandum (Zhao 2002).
\label{fig1}}

\figcaption[fig2.ps]{
The SMA light curves at 1.3 mm observed in the period between
2001 March  and 2002 May for the calibrators, OV 236 and NRAO 530 
(panel a) and
Sgr~A* (panel b). The solid curves in  panel a are
the quadratic fits to the secular variations 
of the flux density of OV 236 and NRAO 530. 
The secular trends of the flux density 
from NRAO 530 and OV 236 are the same as those observed at 3 mm at other 
observatories ({\it e.g.} M. Yun 2002, private communication). 
The densely sampled
radio light curves at 1.3 and 2 cm  observed with the VLA  is
shown in panel c
(Herrnstein \etal~2002).
\label{fig2}}

\figcaption[fig3.ps]{(a) A mean spectrum of Sgr~A* made from the observations 
near the
peak of Flare 1. The flux densities of $6.7\pm1.5$ and
$3.7\pm0.7$  Jy  at  0.87 and 1.3 mm are derived from 
the SMA observations.
The flux density of $1.3\pm0.6$ Jy at 3 mm is  averaged from  the
measurements made with the Nobeyama Millimeter Array (NMA) 
(Tsutsumi, Miyazaki, \& Tsuboi~2002). The data ($1.2\pm0.2$, 
$1.15\pm0.17$, and $1.04\pm0.18$ Jy) at 0.7, 1.3 and 2 cm is derived
from 
the VLA monitoring observations (Herrnstein, {\it et. al}~2002, in preparation). 
The curves represent a resultant spectrum of two power-law components 
(straight lines)
for various sets of spectral indices corresponding
to different emission processes for the sub-millimeter
component (see section 3.3).
(b) A spectrum of Sgr A* made from the observations on  
the same date (2001 June 17) during a minimum of the Sgr A*
light curves (see Fig. 2). The flux density of  $1.1 \pm0.15$
at 1.3 mm is measured with the SMA. The rest of the measurements 
($0.72\pm0.09$, $0.69  \pm 0.07$, and $0.65  \pm 0.06$ Jy 
 at  0.7, 1.3 and 2 cm) are   made from the VLA observations.
\label{fig3}}

\figcaption[fig4.ps]{
The z-transformed discrete cross-correlation function (ZDCF)  of 
the SMA data at 1.3 mm
with the VLA data at 1.3 cm. 
\label{fig4}}

\newpage
\plotone{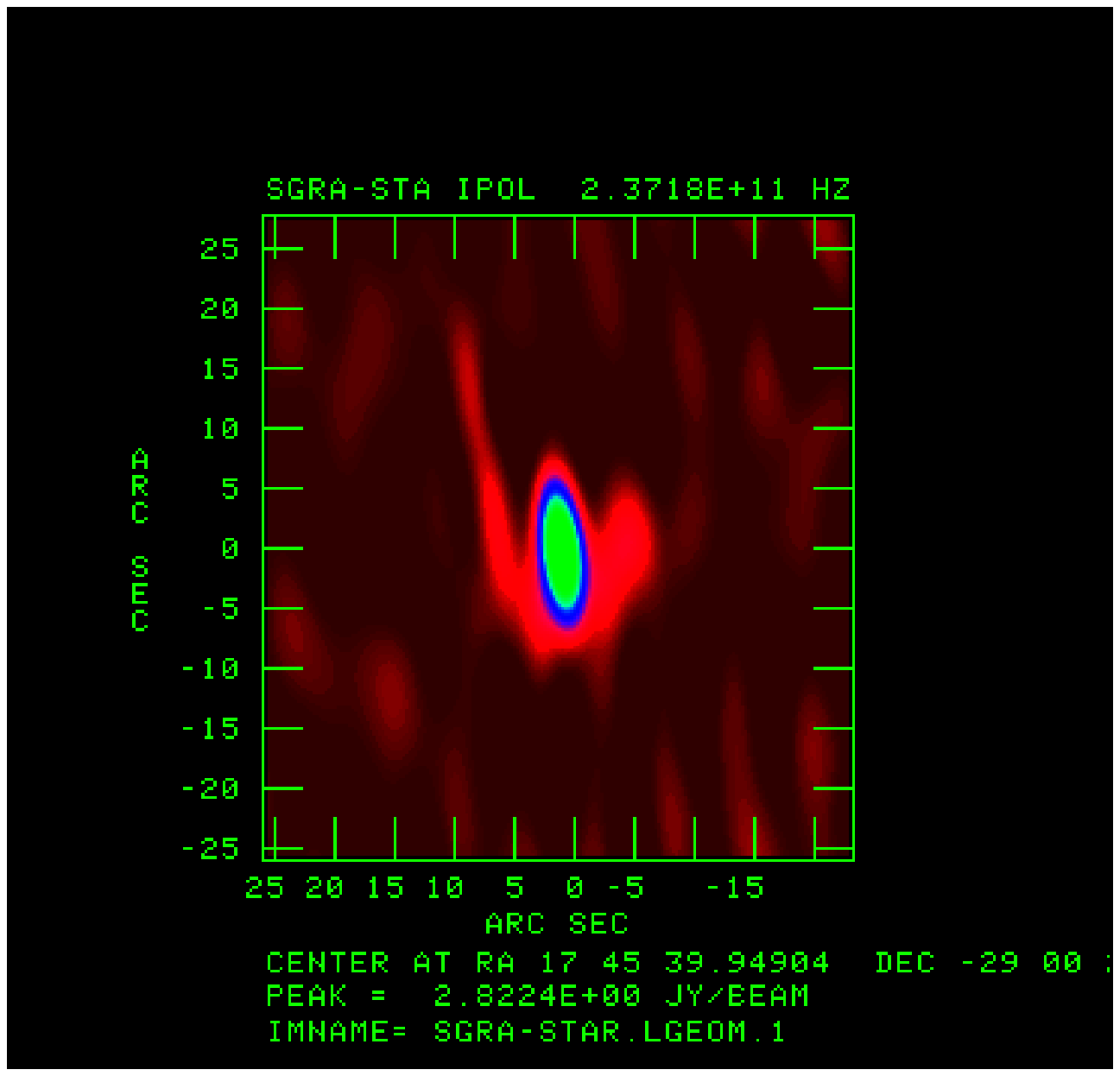}
\newpage
\plotone{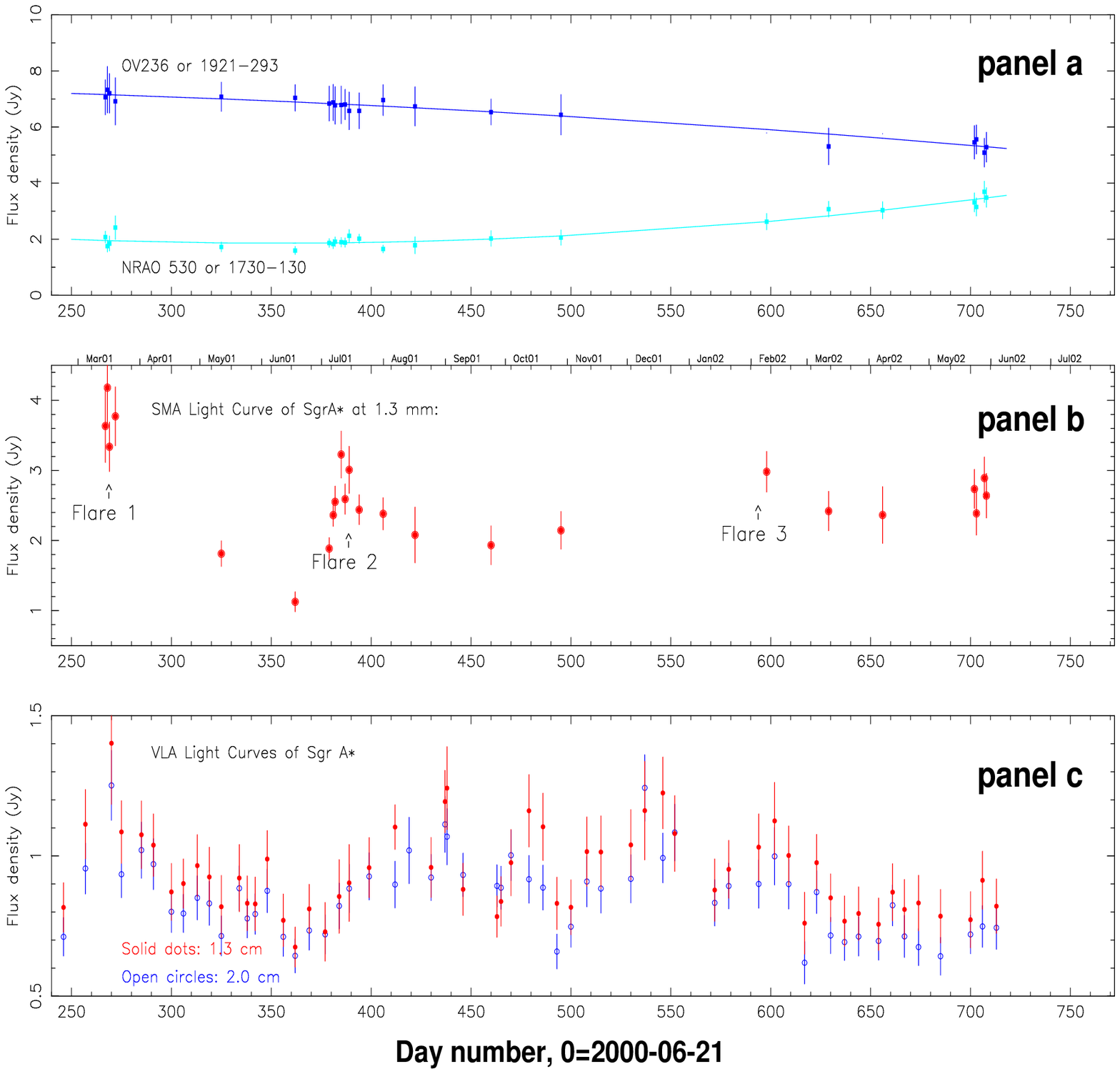}
\newpage
\plotone{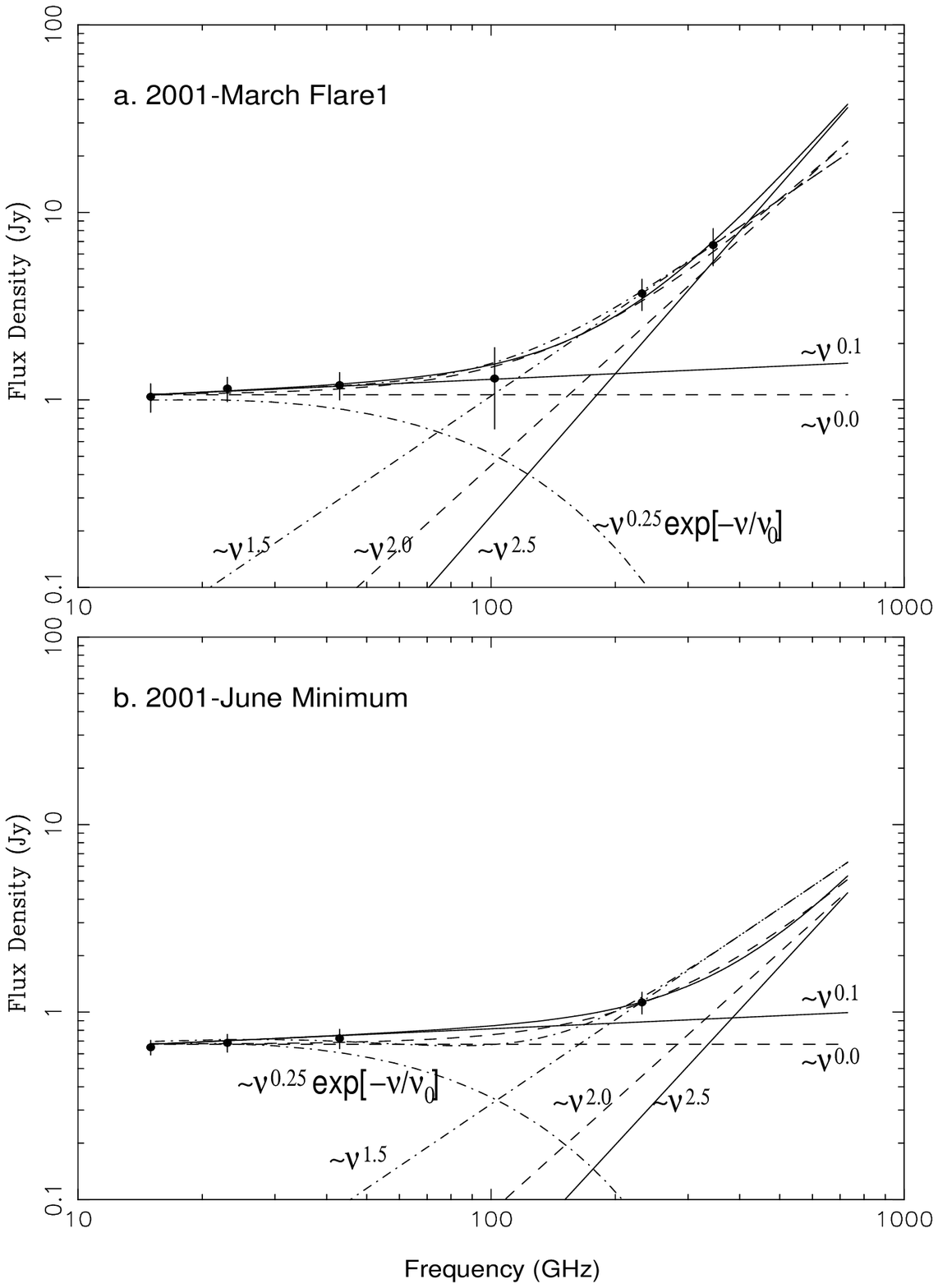}
\newpage
\plotone{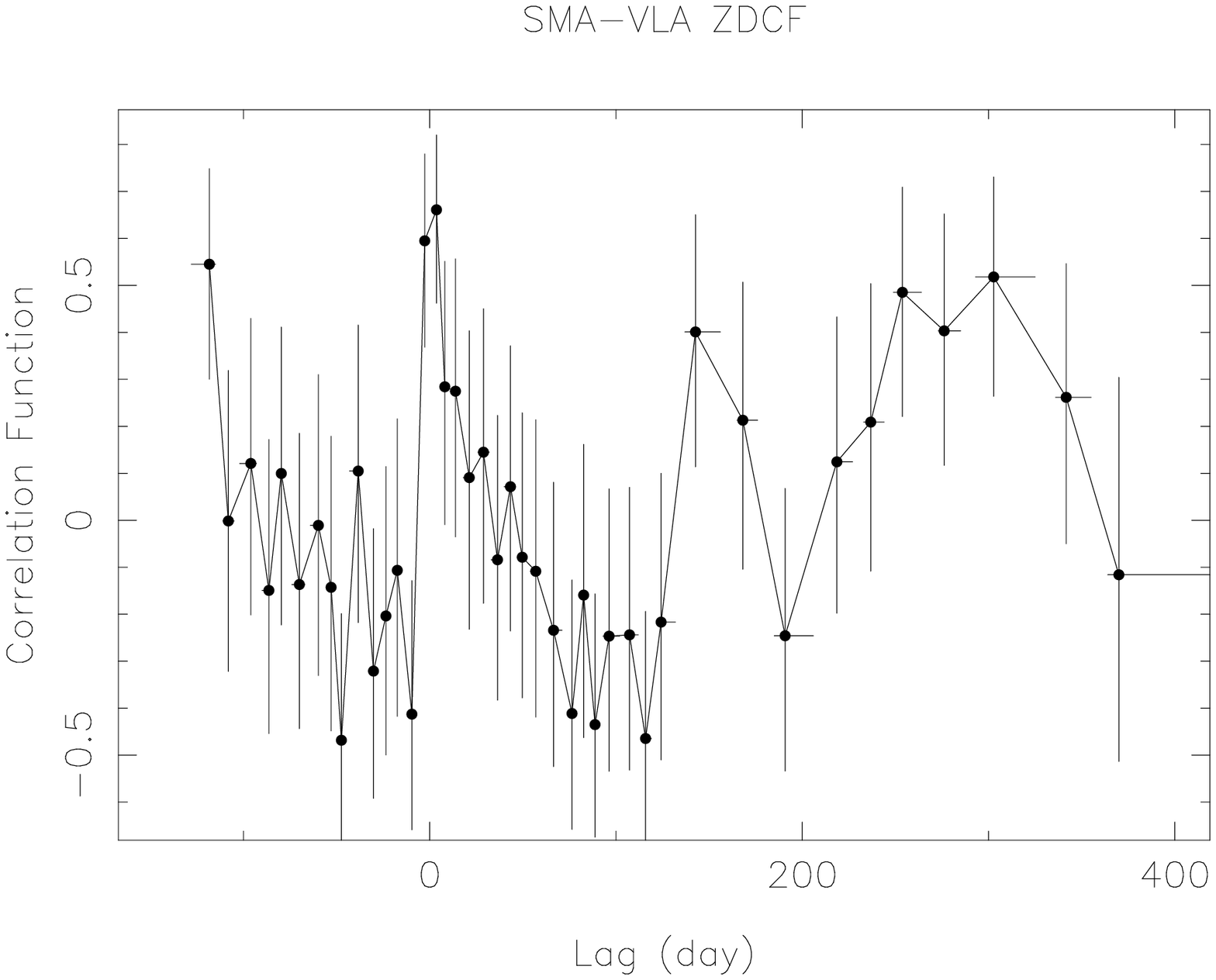}

\end{document}